\documentclass{article}

\usepackage{latexsym}
\usepackage{amsmath,enumerate}
\usepackage[psamsfonts]{amssymb}
\usepackage{amsthm}

\usepackage[mathscr]{eucal}

\newcommand{\pz}{\sqrt{e^{2\phi}+|p|^2}}

\newcommand{\R}{\mathbb{R}}

\newtheorem{Theorem}{Theorem}
\newtheorem{Proposition}{Proposition}
\newtheorem{Lemma}{Lemma}
\newtheorem{Remark}{Remark}

\title{Virial inequalities for steady states\\ in relativistic galactic dynamics\thanks{The
authors have been partially supported by
 Ministerio de Ciencia e Innovaci\'on (Spain), Project MTM2008-05271 and Junta de Andaluc\'{\i}a Project  E--792.} }
\author{Simone Calogero, Juan Calvo,\\ \'Oscar S\'anchez \& Juan Soler\thanks{Departamento de Matem\'atica Aplicada.
Facultad de Ciencias, Universidad de Granada. 
18071 Granada, Spain. ({\tt calogero@ugr.es, juancalvo@ugr.es, ossanche@ugr.es, jsoler@ugr.es})} }
\date { }
\begin{document}
\maketitle
\begin{abstract}{It is well known that steady states of the Vlasov-Poisson system, a widely used model in non-relativistic galactic dynamics, have negative energy. In this paper we derive the analogous property for two relativistic generalizations of the Vlasov-Poisson system: The Nordstr\"om-Vlasov system and the Einstein-Vlasov system. In the first case we show that the energy of steady states is bounded by their total rest mass; in the second case, where  we also assume spherical symmetry, we prove an inequality  which involves not only the energy and the rest mass, but also the central redshift. In both cases the proof makes use of integral inequalities satisfied by time depedent solutions and which are derived using the vector fields multipliers method.}
\end{abstract}
{\bf AMS classification (2000):} 35B05, 82B40.


\section{Introduction and main results}\label{intro}
A widely used model in astrophysics for the dynamics of the stars of a galaxy is the Vlasov-Poisson system~\cite{galactic}. This model is justified when collisions among the stars and external forces are neglected. Moreover, being a non-relativistic model, the Vlasov-Poisson system ceases to be valid when the stars move with large velocities (of the order of the speed of light) or in the presence of very massive galaxies, since then relativistic effects become important. Typical relativistic effects are the redshift of the luminous signals emitted by the galaxy and the formation of black holes. The model which is currentely believed to represent the physically correct relativistic generalization of the Vlasov-Poisson system is the Einstein-Vlasov system~\cite{And}, where Poisson's equation is substituted by Einstein's equations of General Relativity. 
Another relativistic generalization of Vlasov-Poisson is the Nordstr\"om-Vlasov system~\cite{Simone2003}. This model relies on the same geometric interpretation of gravity as for the Einstein-Vlasov system. Although it is not physically correct, the Nordstr\"om-Vlasov system is mathematically interesting since it already captures some of the technical and conceptual new difficulties that are encountered when studying a relativistic (Lorentz invariant) system.

In this paper we want to investigate the mass-energy bounds (virial inequalities) required for the existence of steady states to the relativistic models. It is well known in fact that static solutions of the Vlasov-Poisson system, which correspond to equilibrium configurations of the galaxy, have negative energy. The same property cannot of course be true for the relativistic models, since the energy in the latter case is always positive. 
Before considering the relativistic models in detail we present the role of the virial identities in the case of steady states to  the Vlasov-Poisson system.

\subsection{The classical case: Vlasov-Poisson system}\label{VPsection}
 Let $f=f(t,x,p)$ be the distribution function in phase space for an ensemble of unit mass particles, where $f\geq 0$, $t\in\R$, $x\in\R^3$ and $p\in\R^3$. In the physics of gravitational systems, the particles stand for the stars of a galaxy. 
In geometric units, i.e., $4\pi G=1$, where $G$ is Newton's gravitational constant, the gravitational potential $U=U(t,x)$ generated by the galaxy solves the Poisson equation
\begin{subequations}\label{VPsystem}
\begin{equation}\label{poisson}
\Delta_x U=\rho\:,\quad\lim_{|x|\to\infty}U=0\:,\ \forall\,t\in\R\:,
\end{equation}
where 
\begin{equation}\label{rhodef}
\rho=\int_{\R^3} f\,dp\:
\end{equation}
is the mass density of the galaxy and  the boundary condition  at infinity means that the galaxy is isolated. The assumption that the stars interact only by gravity leads to the Vlasov equation:
\begin{equation}\label{vlasovVP}
\partial_t f +p\cdot\nabla_x f -\nabla _xU\cdot\nabla_p f=0\:.
\end{equation}
\end{subequations}
The system~\eqref{VPsystem} is the Vlasov-Poisson system. 
The energy $H$ and the mass $M$ of a solution are given by 
$$
H=\frac{1}{2}\int_{\R^3}\int_{\R^3}|p|^2f\,dp\,dx-\frac{1}{2}\int_{\R^3}|\nabla_x U|^2dx\:,\quad M=\int_{\R^3}\int_{\R^3}f\,dp\,dx
$$
and are conserved quantities. Likewise, the total linear momentum $Q$ and angular momentum $L$,
$$
Q=\int_{\R^3}\int_{\R^3} p\,f\,dp\,dx\:,\qquad L=\int_{\R^3}\int_{\R^3}x\wedge p\,f\,dp\,dx\:,
$$
are conserved quantities.
The invariance of Vlasov-Poisson by (time dependent) Galilean transformations is the property that, given $u\in\R^3$ and the transformation of coordinates
$$
\mathcal{G}_u:\quad t'=t\:,\quad x'=x-ut\:,\quad p'=p-u\:,
$$
then $f_u(t,x,p)=f(t',x',p')$ and $U_u(t,x)=U(t',x')$ solve the system~\eqref{VPsystem} if and only if $(f,U)$ does. Note that $Q$ can be made to vanish by a Galilean transformation with velocity $u=Q/M$; the resulting reference frame is at rest with respect to the center of mass of the distribution, which is defined as
$c_\rho(t)=M^{-1}\int_{\R^3} x\,\rho\,dx\:$.

A galaxy in equilibrium is described by steady states solutions of the Vlasov-Poisson system. We distinguish between two types of steady states: Static solutions and traveling steady states. The formers are defined as time independent solutions of the Vlasov-Poisson system~\eqref{VPsystem} and have total momentum $Q=0$. A solution $f$ is a traveling steady state (with total momentum $Q\neq 0$) if $f\circ\mathcal{G}_u$, where $u=Q/M$, is a time independent solution of the Vlasov-Poisson system (i.e., a static solution). Our interest on traveling steady states is motivated by the fact that their energy provides a lower limit for the energy of totally dispersive solutions, see~\cite{dispersion}. Moreover, the non-linear stability theorems proved for the Vlasov-Poisson system consider the traveling steady states as possible perturbations of a static equilibria, see~\cite{CSS,ReinLibro,SS} and references therein.  

 A fundamental property shared by all static solutions of the Vlasov-Poisson system is that of having negative energy. The proof goes as follows. Any sufficiently regular solution of the Vlasov-Poisson system satisfies the dilation identity:
\[
\frac{d}{dt} \int_{\R^3}\int_{\R^3}x\cdot p\, f\,dp\,dx=H+E_{\mathrm{kin}}\:,\quad E_{\mathrm{kin}}=\frac{1}{2}\int_{\R^3}\int_{\R^3}|p|^2f\,dp\,dx\:,
\]
as it follows by direct computation.
If $f$ is a static solution, then the previous identity implies the {\it virial} relation $H=-E_{\mathrm{kin}}$,  which yields that 
\begin{equation}\label{negativeEforVP}
H<0\:,\quad\text{for static solutions of the Vlasov-Poisson system.}
\end{equation} 
For traveling steady states, we just apply to~\eqref{negativeEforVP} a Galilean transformation with $u=(M)^{-1}Q$ and we obtain
\begin{equation}\label{negativeEforVP2}
H<\frac{|Q|^2}{2M}\:,\quad\text{for traveling steady states of the Vlasov-Poisson system.}
\end{equation} 
Our purpose in this work is to extend these fundamental inequalities to the relativistic case.
For more information on the Vlasov-Poisson system, we refer to~\cite{dispersion,DSS,GS,ReinLibro}. 

\subsection{Main results for relativistic models}

In the case of the Nordstr\"om-Vlasov system, which will be presented in Section~\ref{NVcase}, the generalization of (\ref{negativeEforVP}) is that the energy of regular steady states is bounded by their mass, i.e.  
\begin{equation}\label{mainresultNV}
H \leq M\: .
\end{equation}  
Furthermore, the counterpart to (\ref{negativeEforVP2}) for traveling steady states is 
$$ \sqrt{H^2 - |Q|^2} \leq M\: .$$
(Of course, the invariants $H$, $M$, $Q$ have to be redefined in an appropriate  way for this new system, see Section~\ref{NVcase}). 
Moreover the equality sign could only hold for steady states with unbounded support. 
We remark that in the case of the Vlasov-Poisson system the supremum of the steady states energy coincide with the infimum energy of totally dispersive time dependent solutions, see~\cite{dispersion}. The analogous statement for the Nordstr\"om-Vlasov system is currently not known, due to the difficulties in defining a Lorentz invariant concept of total dispersion. We also remark that the bound $H<M$, which holds for all regular and compactly supported static solutions of the Nordstr\"om-Vlasov system, is crucial in the proof of orbit stability of the polytropic steady states established in~\cite{CSS}.

\indent For the spherically symmetric Einstein-Vlasov system, analyzed in Section~\ref{EVcase}, we derive an inequality that involves not only the energy (ADM mass, $H$) and the mass (rest mass, $M$) of the steady state, but also the central redshift $Z_ c$:
\begin{equation}\label{mainresultEV}
Z_c \geq \left|\frac{M}{H}-1\right|.
\end{equation}
The metric of the space-time for spherically symmetric static solutions of the Einstein-Vlasov system is determined, following the notation in Section~\ref{EVcase}, by two functions
$\lambda(r) \geq 0$ and $\mu(r) \leq 0$ of the radial variable.  The central redshift is defined in terms of the second one at the origin by $Z_c:=e^{-\mu(0)}-1$. It is the redshift of a photon emitted from the center of the galaxy. The estimate~\eqref{mainresultEV} can thus be seen as an upper bound for $\mu(0)$. Similarly, the celebrated Buchdahl's inequality in General Relativity~\cite{W} can be seen as an upper bound on the metric component $\lambda(r)$ for spherically symmetric steady states of the Einstein-matter equations. A quite general version of the Buchdahl inequality was proved recently in~\cite{And2} and reads
\begin{equation}\label{BOU}
\sup_{r\geq 0} \left(1 - e^{-2\lambda(r)}\right)\leq \frac89\:,\ \text{or equivalently }\ 
\sup_{r \geq 0} \lambda(r) \leq \ln{3}\: .
\end{equation}
For static shells the Buchdahl inequality is equivalent to a lower bound for the external radius. We will show that estimate 
\eqref{mainresultEV} leads to an upper bound on the internal radius.
We refer to~\cite{AnRe2007} for an analitical/numerical investigation of the Buchdahl inequality in the context of the spherically symmetric Einstein-Vlasov system. 

We do not know whether, as for the Vlasov-Poisson and the Nordstr\"om-Vlasov system, the inequality~\eqref{mainresultEV} could also be related to the problem of stability of spherically symmetric static solutions. This is a difficult question to answer, since the stability problem for the Einstein-Vlasov system is still poorly understood. However it is worth noticing that heuristic and numerical studies \cite{Z1,Z2, ST} indicate that the regime of stability of compact galaxies is indeed characterized by the central redshift 
and the fractional binding energy (defined as $1-H/M$). Moreover it was conjectured
that the binding energy maximum along a steady state sequence signals the 
onset of instability. There are several numerical studies on the problem of stability for the spherically Einstein-Vlasov system; we refer to~\cite{AnRe2006,AnRe2007,Rein1998}.

A last basic comment on~\eqref{mainresultEV} is that, as opposed to the inequalities that hold for steady states of the Vlasov-Poisson and the Nordstr\"om-Vlasov system, the bound \eqref{mainresultEV} contains a quantity, the central redshift, which is not preserved along time dependent solutions. It is therefore not clear whether one can interpret~\eqref{mainresultEV}  as the exact analog of the mass-energy inequalities for the steady states of the Vlasov-Poisson and Nordstr\"om-Vlasov system. 

We conclude this Introduction with a brief explanation on how we prove our main results. As a first step we employ the vector fields multipliers method to the local conservation laws for the Nordstr\"om-Vlasov and the Einstein-Vlasov system to establish a virial identity which has to be satisfied by all time dependent solutions. These identities are of independent interest and could be useful to derive space-time (Morawetz type) estimates for the evolution problem. The virial identities restricted to time independent solutions give rise, after applying some simple bounds on the moments of the distribution $f$, to the virial inequalities~\eqref{mainresultNV}-\eqref{mainresultEV}.

\section{The Nordstr\"om-Vlasov case}\label{NVcase}
We write the Nordstr\"om-Vlasov system in the formulation used in \cite{CSS}:
\begin{subequations}\label{NVsystem}
\begin{equation} \label{vlasovt}
\partial_t f +\frac{p}{\pz}\cdot\nabla_x f- 
\nabla_x\left(\pz\right)\cdot\nabla_p f = 0\:,
\end{equation}
\begin{equation}\label{wavet}
\partial_t^2\phi-\Delta_x\phi = - e^{2\phi}\int_{\R^3} 
f\,\frac{dp}{\pz}\:.
\end{equation}
\end{subequations}
Here $f=f(t,x,p)\geq 0$ and  $\phi=\phi(t,x)$. The physical interpretation of a solution $(f,\phi)$ is the following: The space-time is the Lorentzian manifold $(\R^4, g=e^{2\phi}\eta)$, where $\eta$ is the Minkowski metric, whereas $f$ is the kinetic distribution function of particles (the stars of a galaxy)  moving along the geodesic curves of the metric $g$. The motion along geodesics reflects the condition that gravity is the only interaction among the particles. The system has been written in units such that $4\pi G=c=1$, where $G$ is Newton's gravitational constant and $c$ the speed of light. If the latter is restored in the equations, in the limit $c\to\infty$ one recovers the Vlasov-Poisson system in the gravitational case. For a proof of the latter statement and general information on the Nordstr\"om-Vlasov system we refer to~\cite{CL}. The global regularity of solutions is studied in \cite{Bostan,CR}.

The local energy, momentum and stress tensor of a solution $(f,\phi)$ of~\eqref{NVsystem} are defined respectively as ($i,j=1,2,3$)
\begin{subequations}
\begin{align*}\label{localfunctions}
&h(t,x)=\int_{\R^3} 
\sqrt{e^{2\phi}+|p|^2}\,f\,dp+\frac{1}{2}(\partial_t\phi)^2+\frac{1}{2}|\nabla_x\phi|^2,\\
&q_i(t,x)=\int_{\R^3} p_if\,dp-\partial_t\phi\,\partial_i\phi\,,\\
&\tau_{ij}(t,x)=\int_{\R^3} \frac{p_i\,p_j}{\sqrt{e^{2\phi}+|p|^2}}\,f\,dp+\partial_i\phi\,\partial_j\phi+\frac{1}{2}\delta_{ij}\left[(\partial_t\phi)^2-|\nabla_x\phi|^2\right],
\end{align*}
\end{subequations}
where $\partial_i$ denotes the partial derivative along $x^i$.
These quantities are
related by the conservation laws
\begin{equation}\label{localcons}
\partial_th+\nabla_x\cdot q=0\:,\quad\partial_t q_i+\partial_j\tau_{ij}=0\:,
\end{equation}
the sum over repeated indexes being understood.
Upon integration, the previous identities lead to the conservation of 
the total energy and of the total momentum:
\[
H(t)=\int_{\R^3} h(t,x)\,dx=constant\:,\quad Q(t)=\int_{\R^3} q(t,x)\,dx=constant\:.
\]
Moreover, solutions of the Nordstr\"om-Vlasov system satisfy the 
conservation of the total rest mass:
\[
M(t)=\int_{\R^3} \rho(t,x)\,dx=constant\:,
\]
which is obtained by integrating the local rest mass conservation law
\begin{equation}\label{conteq}
\partial_t\rho+\nabla_x\cdot j=0\:,\quad \rho=\int_{\R^3} f\,dp\:,\quad 
j=\int_{\R^3} \frac{p}{\sqrt{e^{2\phi}+|p|^2}}\,f\,dp\:.
\end{equation}

The system~\eqref{NVsystem} satisfies the fundamental property of Lorentz invariance. Precisely, let $(t',x')$ be a system of coordinates in Minkowski space obtained from $(t,x)$ by a Lorentz boost, that is
\[
t'=u_0t- u\cdot x\:,\quad x'=x-u\,t+\frac{u_0-1}{|u|^2}(u\cdot x)u\:,
\]
where $u$ is a fixed vector in $\R^3$ and $u_0=\sqrt{1+|u|^2}$. The inverse Lorentz transformation is obtained by exchanging $u$ with $-u$, that is 
\begin{equation}\label{inverselorentz1}
t=u_0t'+ u\cdot x'\:,\quad x=x'+u\,t'+\frac{u_0-1}{|u|^2}(u\cdot x')u\:,
\end{equation}
which we shorten by $(t,x)=L_u(t',x')$.
Define the field $\phi_u$ in the new coordinates by
\[
\phi_u(t',x')=\phi\circ L_u\,(t',x')\:.
\]
Introduce the new momentum variable
\[
p'=p-u\sqrt{e^{2\phi(t,x)}+|p|^2}+\frac{u_0-1}{|u|^2}(u\cdot p)u
\]
or, inverting,
\begin{equation}\label{inverselorentz2}
p=p'+u\sqrt{e^{2\phi(t,x)}+|p'|^2}+\frac{u_0-1}{|u|^2}(u\cdot p')u\:.
\end{equation}
We shall write $(t,x,p)=\mathcal{L}_u(t',x',p')$ to shorten the set of transformations (\ref{inverselorentz1})-(\ref{inverselorentz2}).
Finally, define the distribution function in the new variables as
\[
f_u(t',x',p')=f\circ\mathcal{L}_u\,(t',x',p')\:.
\]
In the language of special relativity, one says that $f$ and $\phi$ transform like scalar functions under Lorentz transformations. The Lorentz invariance of the Nordstr\"om-Vlasov system means that the pair $(f,\phi)$ solves the system~\eqref{NVsystem} in the coordinates $(t,x,p)$ if and only if $(f_u,\phi_u)$ satisfies the same system in the coordinates $(t',x',p')$. Thus, in particular, also the mass-energy-momentum of $(f_u,\phi_u)$ is conserved along the time evolution,
\[
M[f_u]=constant\:,\quad H[f_u,\phi_u]=constant\:, \quad Q[f_u,\phi_u]=constant\: .
\] 
We shall need the relation between the mass-energy-momentum of $(f,\phi)$ and of $(f_u,\phi_u)$, which is derived the following lemma\footnote{In the language of special relativity, the lemma establishes that $M$ transforms like a scalar function, whereas the quadruple $(H,Q)$ transforms like a four-vector under Lorentz transformations.}.
\begin{Lemma}\label{transformation}
For all $u\in\R^3$, 
\begin{subequations}\label{transformations}
\begin{align}
& M[f_u]=M[f]\:, \label{trans1} \\
& H[f_u,\phi_u]=\sqrt{1+|u|^2}\,H[f,\phi]-Q[f,\phi]\cdot u\:,  \label{trans2} \\
& Q[f_u,\phi_u] = Q[f,\phi] - H[f,\phi] u + \frac{u_0 -1}{|u|^2} (u \cdot Q[f,\phi]) u\:.  \label{trans3} 
\end{align}
\end{subequations}
\end{Lemma}   
\begin{proof}
Since the mass-energy-momentum of both pairs $(f,\phi)$ and $(f_u,\phi_u)$ is conserved, it is sufficient to prove the relations~\eqref{transformations}  for the initial value of $M(u):=M[f_u]$, $H(u):=H[f_u,\phi_u]$ and $Q(u):=Q[f_u,\phi_u]$. We restrict ourselves to prove the invariance of the total mass, the proof for the other transformations being similar.
We shall need that, by (\ref{inverselorentz2}),
\[
\sqrt{e^{2\phi(t,x)}+|p'|^2}=u_0\sqrt{e^{2\phi(t,x)}+|p|^2}-u\cdot p
\]
or, inverting,
\begin{equation}\label{inverselorentz3}
\sqrt{e^{2\phi(t,x)}+|p|^2}=u_0\sqrt{e^{2\phi(t,x)}+|p'|^2}+u\cdot p'.
\end{equation}
In order to prove (\ref{trans1}) we write
\begin{align*}
M(u)&=\int_{\R^3}\int_{\R^3} f_u(0,x',p')\,dx'\,dp'=\int_{\R^3}\int_{\R^3} f\circ\mathcal{L}_u(0,x',p')\,dx'\,dp'\\
&=\int_{\R^3}\int_{\R^3} f\left(u\cdot x',x'+\frac{u_0-1}{|u|^2}(u\cdot x')u,p'+u\sqrt{e^{2\phi_u(0,x')}+|p'|^2}+\frac{u_0-1}{|u|^2}(u\cdot p')u\right)dx'\,dp'.
\end{align*}
Next we make the change of variable
\[
x=x'+\frac{u_0-1}{|u|^2}(u\cdot x')u\:,\quad p=p'+u\sqrt{e^{2\phi_u(0,x')}+|p'|^2}+\frac{u_0-1}{|u|^2}(u\cdot p')u\:.
\]
The Jacobian of this transformation is given by
\[
J=\frac{\widehat{u}\cdot p'+\sqrt{e^{2\phi_u(0,x')}+|p'|^2}}{\sqrt{e^{2\phi_u(0,x')}+|p'|^2}}\:,
\]
where $\widehat{u}=u/u_0$.
Using (\ref{inverselorentz2}) and (\ref{inverselorentz3}) we obtain
\[
J=\left(1-\frac{\widehat{u}\cdot p}{\sqrt{e^{2\phi_u(0,x')}+|p|^2}}\right)^{-1}\:.
\]
Since the volume measure  transforms as $dx'dp'=J^{-1}dx\,dp$, we obtain
\begin{align*}
M(u)&=\int_{\R^3}\int_{\R^3} f(\widehat{u}\cdot x,x,p)\left(1-\frac{\widehat{u}\cdot p}{\sqrt{e^{2\phi(\widehat{u}\cdot x,x)}+|p|^2}}\right)\,dx\,dp\\
&=\int_{\R^3}\left(\rho(\widehat{u}\cdot x,x)-\widehat{u}\cdot j(\widehat{u}\cdot x,x)\right)\,dx\:,
\end{align*}
where $\rho$ and $j$ are defined by (\ref{conteq}). Taking the partial derivative $\partial_{u_i}$ of the previous expression we get
\begin{align*}
\partial_{u_i}M(u)&=\int_{\R^3} \left(\partial_t\rho\,\partial_{u_i}(\widehat{u}\cdot x)-(\partial_{u_i}\widehat{u}_k)j_k-\widehat{u}_k\partial_t j_k\partial_{u_i}(\widehat{u}\cdot x)\right)(\widehat{u}\cdot x,x)\,dx\\
&=\int_{\R^3} \left(-\partial_{u_i}(\widehat{u}\cdot x)(\partial_{x_k}j_k+\widehat{u}_k\partial_t j_k)-(\partial_{u_i}\widehat{u}_k)j_k\right)(\widehat{u}\cdot x,x)\,dx\\
&=-\int_{\R^3}\left(\partial_{u_i}(\widehat{u}\cdot x)\partial_{x_k}[j_k(\widehat{u}\cdot x,x)]- (\partial_{u_i}\widehat{u}_k)j_k(\widehat{u}\cdot x,x)\right)\,dx\\
&=0\:,
\end{align*}
where we used the continuity equation (\ref{conteq}) to pass from the first to the second line and integration by parts to pass from the third to the last line. Thus we obtained that $\nabla_u M(u)=0$, i.e., $M(u)=M(0)$, which yields the claim on the invariance of the total rest mass.
\end{proof}

\begin{Remark}\textnormal{ According to the transformation law of the total momentum $Q$, the Lorentz transformation that makes $Q$ to vanish, i.e. that moves the reference frame to the center of mass system\footnote{To be more precise, this is called the {\it center of momentum system}. In relativity there is no general acceptance on the concept of center of mass.}, is the transformation $\mathcal{L}_u$ with $u=Q/\sqrt{H^2 - |Q|^2}$. The  energy of the transformed solution is $\sqrt{H^2 - |Q|^2}$.} 
\end{Remark}

\subsection{Virial identities for time dependent solutions}
The conservation laws~\eqref{localcons} can be expressed in a more coincise form as 
\begin{equation}\label{consTNordstrom}
\partial_\mu T^{\mu}_{\ \nu}=0\:,\qquad \mu,\nu=0,\dots,4\:,\ x^0=t\:,
\end{equation} 
where $T_{\mu\nu}$ is the stress-energy tensor, whose components are given by
\[
T_{00}=-h\:,\ T_{0i}=-q_i\:,\ T_{ij}=\tau_{ij}\:.
\]
Indexes are raised and lowered with Minkowski's metric $\eta_{\mu\nu}=\mathrm{diag}(-1,1,1,1)$. Upon multiplying the conservation law~\eqref{consTNordstrom}  by a vector field $\xi^\mu=\xi^\mu(t,x)$, integrating on a compact spacetime region $\Omega$ with piecewise differentiable boundary $\partial\Omega$ and applying the divergence theorem we obtain the integral identity
\begin{equation}\label{generalintegralidentity}
\int_{\partial\Omega}T^\mu_{\ \nu}\xi^\nu n_\mu d\sigma=\int_{\Omega} T^\mu_{\ \nu} \partial_\mu\xi^\nu dtdx\:,
\end{equation}
where $n_\mu$ denotes the exterior normal vector field to the boundary $\partial\Omega$ and $d\sigma$ the invariant volume measure thereon. The identities obtained from~\eqref{generalintegralidentity} upon a specific choice of the vector field multiplier go under the general name of {\it virial identities}. We prove here one that applies to {\it regular asymptotically flat solutions}. By this we mean that $f\in C^1$, $\phi\in C^2\cap L^2$, the mass and energy are finite and
\begin{equation}\label{decay}
\lim_{R\to\infty}\int_{S(R)}h(t,x) dS_R = 0\:,\ \forall\, t\in\R\:.
\end{equation}
By $\omega$ we shall denote the outward unit normal to $S(R)=\{x:|x|=R\}$, and $dS_R$ stands for the invariant volume measure on $S(R)$. Moreover we denote by $\chi(r)$, $r>0$, a function that satisfies:
\begin{equation}\label{propchi1}
\chi\in C^2\:,\quad \chi'\in L^{\infty}\:,\quad\frac{\chi}{r}\in C^2\cap L^\infty\:. 
\end{equation}
\begin{Lemma}\label{integralidentity} Let
\[
\mathcal{I}(t)=\int_{\R^3}\chi(r)\left(q\cdot\omega-r^{-1}\phi\,\partial_t\phi\right)dx\:,\ r=|x|\:.
\]
For all regular asymptotically flat solutions of~\eqref{NVsystem} the following identity holds:
\begin{align}\label{mainidentity}
\frac{d\mathcal{I}}{dt}=&\int_{\R^3}\chi'\,h\,dx+\int_{\R^3}\frac{\chi}{r}e^{2\phi}(\phi-1)\int_{\R^3}\frac{f}{\sqrt{e^{2\phi}+|p|^2}}\,dp\,dx\nonumber\\
&+\int_{\R^3}\left(\frac{\chi}{r}-\chi'\right)\left[|\omega\wedge\nabla_x\phi|^2+\int_{\R^3}\frac{|\omega\wedge p|^2 + e^{2 \phi}}{\sqrt{e^{2\phi}+|p|^2}}\,f\,dp\right]dx\nonumber\\
&-\frac{1}{2}\int_{\R^3}\frac{\chi''}{r}\phi^2dx\:.
\end{align}
\end{Lemma}
\begin{proof}
 In~\eqref{generalintegralidentity} we use $\Omega=[0,T]\times B(R)$, where $B(R)=\{x:|x|\leq R\}$ and
\[
\xi^\mu\: :\: \xi^0=0\:,\ \xi^i=\chi(r)\omega^i\, .
\]
We obtain
\begin{align}\label{firststep}
\left[\int_{B(R)}\chi(r)q\cdot\omega\,dx\right]_0^T=&\int_0^T\int_{S(R)}\chi(r)\tau_{ij}\omega^i\omega^jdS_Rdt\nonumber\\
&+\int_0^T\int_{B(R)}\left[ \left( \chi'-\frac{\chi}{r}\right)\tau_{ij}\omega^i\omega^j+\frac{\chi}{r}\delta^{ij}\tau_{ij}\right]dx\,dt\:,
\end{align}
where for any function $g(t)$ we denote $[g(t)]_0^T=g(T)-g(0)$.
Using the bound $|\tau_{ij}\omega^i\omega^j|\leq 3h$ and~\eqref{decay} we get
\[
\left|\int_{S(R)}\chi(r)\tau_ {ij}\omega^i\omega^jdS_R\right|\leq 3\|\chi\|_\infty\int_{S(R)}h\,dS_R\to 0,\ R\to\infty\:.
\]
Then, letting $R\to\infty$ in~\eqref{firststep} we obtain 
\[
\left[\int_{\R^3}\chi(r)q\cdot\omega\,dx\right]_0^T=\int_0^T\int_{\R^3}\left[ \left( \chi'-\frac{\chi}{r}\right)\tau_{ij}\omega^i\omega^j+\frac{\chi}{r}\delta^{ij}\tau_{ij}\right]dx\,dt\:,
\]
whence
$$
\frac{d}{dt}\int_{\R^3}\chi(r)q\cdot\omega\,dx=\int_{\R^3}\left[ \left( \chi'-\frac{\chi}{r}\right)\tau_{ij}\omega^i\omega^j+\frac{\chi}{r}\delta^{ij}\tau_{ij}\right]dx\:.
$$
We compute 
$$
  \delta^{ij}\tau_{ij} = \int_{\R^3} \frac{|p|^2 f \, dp}{\sqrt{e^{2 \phi}+|p|^2}} + \frac{3}{2}(\partial_t\phi)^2 - \frac{1}{2} |\nabla_x \phi|^2
$$
and
$$
   \tau_{ij} \omega^i \omega^j = \int_{\R^3} \frac{(\omega \cdot p)^2 f \, dp}{\sqrt{e^{2 \phi}+|p|^2}}  + (\omega \cdot \nabla_x \phi)^2 + \frac{1}{2} [(\partial_t\phi)^2 - |\nabla_x \phi|^2]\:.
$$
Hence
\begin{align*}
\frac{d}{dt}\int_{\R^3}\chi(r)q\cdot\omega\,dx =& \int_{\R^3} \frac{\chi}{r} \left[ \int_{\R^3} \frac{|p|^2 f \, dp}{\sqrt{e^{2 \phi}+ |p|^2}} + \frac{3}{2} (\partial_t\phi)^2 - \frac{1}{2} |\nabla_x \phi|^2\right] \, dx\nonumber\\
& + \int_{\R^3} \left(\chi' - \frac{\chi}{r} \right) \left[\int_{\R^3} \frac{(\omega\cdot p)^2 f \, dp}{\sqrt{e^{2 \phi}+|p|^2}} + (\omega\cdot \nabla_x \phi)^2 + \frac{1}{2}[(\partial_t\phi)^2 - |\nabla_x \phi|^2) \right] \, dx\:.\nonumber\\
\end{align*}
Using that $|\omega \wedge y|^2 = |y|^2 - |\omega \cdot y|^2$, for all vectors $y\in\R^3$, we can rewrite the previous equation as
\begin{align}\label{temp1}
\frac{d}{dt}\int_{\R^3}\chi(r)q\cdot\omega\,dx=&  \int_{\R^3} \frac{\chi}{r} ((\partial_t\phi)^2 - |\nabla_x \phi|^2)\, dx - \int_{\R^3} \chi' \int_{\R^3} \frac{e^{2\phi} f \, dp}{\sqrt{e^{2 \phi}+|p|^2}}\, dx\nonumber\\
&  + \int_{\R^3} \chi' h \, dx    +\int_{\R^3}\left(\frac{\chi}{r} - \chi' \right) \left(\int_{\R^3} \frac{|\omega \wedge p|^2 f \, dp}{\sqrt{e^{2 \phi}+|p|^2}} +  |\omega\wedge \nabla_x \phi|^2  \right)\, dx\:.
\end{align}
Moreover, using~\eqref{wavet} and integrating by parts twice, we find
\begin{align}\label{temp2}
\frac{d}{dt}\int_{B(R)}\frac{\chi}{r}\,\phi\,\partial_t\phi\,dx
=&\int_{B(R)}\frac{\chi}{r}\left((\partial_t\phi)^2-|\nabla_x\phi|^2-\phi\int_{\R^3} \frac{e^{2\phi}f\,dp}{\sqrt{e^{2\phi}+|p|^2}}\right)dx\nonumber\\
&+\frac{1}{2}\int_{B(R)}\Delta\left(\frac{\chi}{r}\right)\phi^2\,dx+\int_{S(R)}\frac{\chi}{r}\phi\,\omega\cdot\nabla_x\phi\, dS_R\nonumber\\
&-\frac{1}{2}\int_{S(R)}\omega\cdot\nabla\left(\frac{\chi}{r}\right)\phi^2dS_R\:.
\end{align}
Applying the Cauchy-Schwartz inequality, the regularity of the solution and the assumptions on $\chi$, we obtain
\begin{align*}
\left|\int_{S(R)}\frac{\chi}{r}\phi\,\omega\cdot\nabla_x\phi\, dS_R\right|&\leq C \|\phi\|_{L^2(S(R))}\|\nabla_x\phi\|_{L^2(S(R))}\\
&\leq C\sqrt{\int_{S(R)}h(t,x)\,dS_R}\to 0\:,\ R\to\infty\:,
\end{align*}
and
\begin{align*}
\left|\int_{S(R)}\omega\cdot\nabla\left(\frac{\chi}{r}\right)\phi^2dS_R\right|=\frac{1}{R}\int_{S(R)}\left|\chi'-\frac{\chi}{r}\right|\phi^2dS_R\leq \frac{C}{R}\:,
\end{align*}
where $C$ is a constant independent from $R$. Thus taking the limit $R\to\infty$ in ~\eqref{temp2}  we get
\begin{align}\label{temp3}
-\frac{d}{dt}\int_{\R^3}\frac{\chi}{r}\,\phi\,\partial_t\phi\,dx
=&-\int_{\R^3}\frac{\chi}{r}\left((\partial_t\phi)^2-|\nabla_x\phi|^2-\phi\int_{\R^3} \frac{e^{2\phi}f\,dp}{\sqrt{e^{2\phi}+|p|^2}}\right)dx\nonumber\\
&-\frac{1}{2}\int_{\R^3}\Delta\left(\frac{\chi}{r}\right)\phi^2\,dx\:.
\end{align} 
The quantity $\Delta\left(\frac{\chi}{r}\right)$ is nothing but $\frac{\chi''}{r}$.
%
%
The sum of~\eqref{temp1} and~\eqref{temp3} yields the desired result. 
\end{proof}

\subsection{Virial inequalities for steady states}
As in the Vlasov-Poisson case, we distinguish between two types of steady states. Static solutions, which are defined as time independent solutions of the Nordstr\"om-Vlasov system~\eqref{NVsystem}, and traveling steady states, which are defined as solutions $f(t,x,p)$ such that $f\circ\mathcal{L}_u$, where $u=Q/\sqrt{H^2 - |Q|^2}$, is a time independent solution of the Nordstr\"om-Vlasov system (i.e., a static solution). For static solutions one has $Q=0$, whereas $Q\neq 0$ for traveling steady states. 
Note that for static solutions (that vanish at infinity) the field is determined by $f$ through a non-linear Poisson equation. Thus when we refer to a steady state solution we mean simply the distribution function $f$.  
The main goal of this section is to prove the following property of steady states to the system~\eqref{NVsystem}.
\begin{Theorem}\label{massenergysteadystates}
Let $f$ be a static regular asymptotically flat solution of~\eqref{NVsystem}. Then 
\begin{equation}\label{energymassNV}
H\leq M\:.
\end{equation}
Traveling steady states satisfy $\sqrt{H^2-|Q|^2}\leq M$. Moreover, equality in~\eqref{energymassNV} implies that the support of the static solution is unbounded. 
\end{Theorem}
\begin{proof}
The statement on traveling steady states follows by applying the Lorentz transformation $\mathcal{L}_u$ with $u=Q/\sqrt{H^2 - |Q|^2}$ to the inequality for static solutions, thus it suffices to prove the latter. To this purpose
consider a function $\chi$ that, in addition to~\eqref{propchi1}, satisfies
\begin{equation}\label{propchi2}
\frac{\chi}{r}-\chi'\geq 0\:,\quad \chi''\leq 0\:.
\end{equation} 
Next we observe the simple inequality $y-1\geq -e^{-y}$, with equality if and only if $y=0$. 
Using
\begin{equation}\label{bounduse}
\phi-1\geq -e^{-\phi}
\end{equation}
in the identity~\eqref{mainidentity} we obtain
\begin{equation}\label{maininequality}
\frac{d\mathcal{I}}{dt}\geq \int_{\R^3}\left(\chi'h-\frac{\chi}{r}\rho\right)dx\:.
\end{equation}
In particular, for time independent solutions we have
\begin{equation}\label{maininequalitysteadystates}
 \int_{\R^3}\left(\chi'h-\frac{\chi}{r}\rho\right)dx\leq 0\:.
\end{equation}
Let $R>0$ and consider the function $\chi(r)=\chi_R(r)$ given by 
\[
\chi(r)=\left\{\begin{array}{ll} r&\textnormal{for }r\leqslant R\:,\\ 3R-\frac{3R^2}{r}+\frac{R^3}{r^2} &\textnormal{for }r>R\:.\end{array}\right.
\]
This function satisfies the properties~\eqref{propchi1} and~\eqref{propchi2}. The left hand side of~\eqref{maininequalitysteadystates} becomes
\begin{align}\label{secondinequalitysteadystates}
 \int_{\R^3}\left(\chi'h-\frac{\chi}{r}\rho\right)dx&=\int_{B(R)}(h-\rho)\,dx+\int_{B(R)^c}\left(\chi'h-\frac{\chi}{r}\rho\right)\nonumber\\
&\geq  \int_{B(R)}(h-\rho)\,dx-C\left(\int_{B(R)^c}h\,dx+\int_{B(R)^c}\rho\,dx \right)\nonumber\\
&= \int_{B(R)}(h-\rho)\,dx+\varepsilon(R)\:,
\end{align}
where $\varepsilon(R)\to 0$ as $R\to\infty$. Thus, assuming $H>M$, there exists $R_0>0$ such that $\varepsilon(R)<(H-M)/4$ and $\int_{B(R)}(h-\rho)\,dx> (H-M)/2$, for all $R>R_0$, whence 
\[
 \int_{\R^3}\left(\chi'h-\frac{\chi}{r}\rho\right)dx>\frac{1}{4}(H-M)>0\:,
\]
which contradicts~\eqref{maininequalitysteadystates}. This concludes the proof of~\eqref{energymassNV}. To prove that the strict inequality holds for static solutions with compact support, we observe that in the latter case the field $\phi$ never vanishes in the support of $f$ (it is strictly negative) and thus the stronger inequality $\phi-1>e^{-\phi}$ holds instead of~\eqref{bounduse}. Thus also the inequality in~\eqref{maininequalitysteadystates} is strict. Since the last member of~\eqref{secondinequalitysteadystates} goes to zero for $R\to\infty$ when $H=M$, the claim follows.\end{proof}

\begin{Remark}\textnormal{Theorem~\ref{massenergysteadystates} improves a similar result proved in~\cite{CSS} in two aspects. Firstly, in~\cite{CSS} the fact that the strict inequality holds for compactly supported steady states was overlooked. Secondly the result presented here requires less decay than the inequality proved in~\cite{CSS} and therefore applies to more general steady states. In particular, this result allows to remove some technical hypothesis in the stability result obtained in~\cite{CSS}. }
\end{Remark}

\section{The Einstein-Vlasov case}\label{EVcase} 
The spherically symmetric
Einstein-Vlasov system in Schwarzschild coordinates is given by the
following set of equations (in units $G=c=1$):
\begin{equation}\label{vlasov1}
\partial_tf+e^{\mu-\lambda}\frac{v}{\sqrt{1+|v|^2}}\cdot\nabla_xf-\left(\lambda_t\frac{x\cdot
v}{r}+
e^{\mu-\lambda}\mu_r\sqrt{1+|v|^2}\right)\frac{x}{r}\cdot\nabla_vf=0\:,
\end{equation}
\begin{subequations}\label{EVsystem}
\begin{align}
&e^{-2\lambda}(2r\lambda_r-1)+1=8\pi r^2h\:,\label{einstein1}\\
&e^{-2\lambda}(2r\mu_r+1)-1=8\pi r^2p^\mathrm{rad}\:,\label{einstein2}
\end{align}
\begin{equation}
\lambda_t=-4\pi r e^{\lambda+\mu}q\:,\label{einstein3}
\end{equation}
\begin{equation}\label{einstein4}
e^{-2\lambda}\left(\mu_{rr}+(\mu_r-\lambda_r)(\mu_r+\frac{1}{r})\right)-e^{-2\mu}\left(\lambda_{tt}+\lambda_t(\lambda_t-\mu_t)\right)=4\pi p^\mathrm{tan}\;,
\end{equation}
\end{subequations}
where
\begin{align*}
&h(t,r)=\int_{\R^3}\sqrt{1+|v|^2}f dv\;,\quad p^\mathrm{rad}(t,r)=\int_{\R^3}
\left(\frac{x\cdot
v}{r}\right)^2f\frac{dv}{\sqrt{1+|v|^2}}\:,\\
&q(t,r)=\int_{\R^3}\ \frac{x\cdot v}{r} f dv\:,\quad
p^\mathrm{tan}(t,r)=\int_{\R^3}\left|\frac{x\wedge
v}{r}\right|^2f\frac{dv}{\sqrt{1+|v|^2}}\:.
\end{align*}
The functions $p^{\mathrm{rad}}$ and $p^\mathrm{tan}$ are the radial and tangential pressure; $h$ is the energy density and $q$ the local momentum density\footnote{We adopt the same notation as in the previous sections, although this differs from the standard notation for the Einstein-Vlasov system.}. 
As usual, $f\geq 0$ is the distribution function of particles (stars) in 
the phase space in the coordinates $t\in\R,\, x\in\R^3,\, v\in\R^3$. The variable $v$ is not the canonical momentum of the particles, the latter being denoted by $p$ in the previous sections.
The function $f$ is spherically symmetric in the sense that
$f(t,x,v)=f(t,Ax,Av)$, for all $A\in SO(3)$. The symbol $\wedge$
denotes the standard vector product in $\R^3$, and for a function $g=g(t,r)$, $r=|x|$, we denote by $g_t$ and $g_r$ the time and radial derivative, respectively. By abuse of
notation, $g(t,r)=g(t,x)$ for any spherically symmetric function. The functions $\lambda,\,\mu$  determine
the metric of the space-time according to
\begin{equation}\label{metric}
ds^2=-e^{2\mu}dt^2+e^{2\lambda}dr^2+r^2d\omega^2,
\end{equation}
where $d\omega^2$ is the standard line element on the unit sphere.
The system is supplied with the boundary conditions
\begin{equation}\label{boundary}
\lim_{r\to\infty}\lambda(t,r)=\lim_{r\to\infty}\mu(t,r)=\lambda(t,0)=0\:,
\end{equation}
which define the asymptotically flat solutions with a regular
center, and the initial condition
\[
0\leq f(0,x,v)=f^\mathrm{in}(x,v),\ f^\mathrm{in}(Ax,Av)=f^\mathrm{in}(x,v)\:,\ \forall A\in
SO(3)\:.
\]
The reader is referred to \cite[section
1.1]{G} for a detailed derivation of the system. Throughout this paper we assume that $f$ is a regular solution of~\eqref{vlasov1}-\eqref{EVsystem} in the sense defined in \cite{RR}. In particular, $f(t,x,v)$ is $C^1$ and has compact support in $(x,v)$, for $t\in [0,T]$, and for any $T>0$. For regular solutions, the metric coefficients are $C^2$ functions of their arguments. 
We emphasize that the existence and uniqueness of global regular solutions to the Cauchy problem for the system~\eqref{vlasov1}-\eqref{EVsystem} is open for general initial data. We also remark that the equation
\begin{equation}\label{auxiliary}
\lambda_r+\mu_r=4\pi r e^{2\lambda}(h+p^\mathrm{rad})\:,
\end{equation}
follows by (\ref{einstein1})-(\ref{einstein2}); by (\ref{auxiliary}) we have $\lambda_r+\mu_r\geqslant 0$ and so, by (\ref{boundary}),
\begin{equation}\label{auxiliary2}
0\geqslant \lambda+\mu\geqslant \mu(0,t)\:.
\end{equation} 

The ADM mass (or energy) $H$ and the total rest mass $M$ of a solution to the spherically symmetric Einstein-Vlasov system are defined by
\begin{equation}\label{adm}
H=\int_{\R^3}\int_{\R^3}\sqrt{1+|v|^2}\,f\,dv\,dx\:,\quad M=\int_{\R^3}\int_{\R^3} e^{\lambda} f\,dv\,dx
\end{equation}
and are constant for regular solutions\footnote{The other two conserved quantities, the linear momentum $Q$ and angular momentum $L$, considered in the previous sections are identically zero in the present context by spherical symmetry.}.  Related to the ADM mass we have the quasi-local mass, defined by
\begin{equation}\label{quasilocalmass}
m(t,r)=4\pi\int_0^r s^2h(t,s)\,ds= \frac{r}{2}\left(1-e^{-2\lambda}\right),
\end{equation}
where we used~\eqref{einstein1}. 
Thus $\lim_{r \to \infty} m(t,r) = H$.

For later convenience, we recall that the non-zero Christoffel symbols for the metric~\eqref{metric} are given by 
\[
\Gamma_{\ 00}^0 = \mu_t\:, \quad \Gamma_{\ 0a}^0=\mu_r \frac{x_a}{r}\:, \quad \Gamma_{\ ab}^0= e^{2(\lambda - \mu)}\lambda_t \frac{x_ax_b}{r^2}\:,
\]
\[
\Gamma_{\ 00}^a = e^{-2(\lambda - \mu)}\mu_r \frac{x^a}{r} \:, \quad
\Gamma_{\ 0b}^a = \lambda_t\frac{x^ax_b}{r^2}\:,
\]
\[
\Gamma_{\ ab}^c=\lambda_r \frac{x^cx_bx_a}{r^3} + \frac{1-e^{-2 \lambda}}{r}\left(\delta_b^c-\frac{x_bx^c}{r^2} \right)\frac{x_a}{r}\:.
\]
Note also that $|g|=e^{2\lambda + 2 \mu}$ is the determinant of the metric.

The stress-energy tensor $T^{\mu\nu}$ for Vlasov matter in spherical symmetry is given by
\begin{subequations}\label{Tvlasov}
\begin{equation}
T^{00}=e^{-2\mu}h\:,\quad T^{0a}=e^{-\lambda-\mu}q\,\frac{x^a}{r}\:,
\end{equation}
\begin{equation}
T^{ab}=e^{-2\lambda}p^\mathrm{rad}\,\frac{x^ax^b}{r^2}+\frac{1}{2}p^\mathrm{tan}\left(\delta^{ab}-\frac{x^ax^b}{r^2}\right)
\end{equation}
\end{subequations}
and satisfies the conservation law\footnote{The identities~\eqref{consT} are a consequence of the Vlasov equation alone, see~\cite{Eh}.}
\begin{equation}\label{consT}
\nabla_\mu T^{\mu\nu}=0\:.
\end{equation}

\subsection{Virial identities for time dependent solutions}

To begin with we derive an integral identity for the spherically symmetric Einstein-Vlasov system as we did in Lemma~\ref{integralidentity} for the Nordstr\"om-Vlasov system, i.e., using the vector fields multipliers method. Actually, the identity in Lemma~\ref{integralidentityEV} below is valid not only for the Einstein-Vlasov system, but for all matter models in spherical symmetry. This is due to the fact that equation~\eqref{consT}, which is the starting point for deriving the integral identity, must be satisfied by all matter models for compatibility with the Einstein equations.   

Multiplying the conservation law~\eqref{consT}  by a vector field $\xi^\mu$, integrating on a compact spacetime region $\Omega$ with piecewise differentiable boundary $\partial\Omega$ and applying the divergence theorem we obtain the integral identity
\begin{equation}
\label{divergence_theorem}
   \int_{\partial \Omega} J^\mu \eta_\mu \, d\sigma_g = \int_\Omega T^{\mu \nu}\nabla_\mu \xi_\nu \, dg\:,
\end{equation}
where $\eta_\mu$ is the normal covector related to the boundary, $J^\mu = T_{\,\,\ \nu}^\mu \xi^\nu = T^{\mu \nu} \xi_\nu$ is the current associated to the vector field $\xi^\mu$ and $\nabla_\mu \xi_\nu = \partial_\mu \xi_\nu - \Gamma_{\mu \nu}^\sigma \xi_\sigma$ is the covariant derivative of the vector field. Moreover $dg$ is the invariant volume element on the spacetime and $d\sigma_g$ the invariant volume element induced on $\partial\Omega$. 

\begin{Lemma}\label{integralidentityEV}
Assume that $(h,q,p^\mathrm{rad},p^\mathrm{tan})$ satisfy the compatibility condition\footnote{In the case of a perfect fluid, the compatibility condition~\eqref{consT} is the system of Euler equations.}~\eqref{consT}, where $T_{\mu\nu}$ is the stress-energy tensor~\eqref{Tvlasov}. In addition, we assume that $h(t,\cdot), q(t,\cdot)$, $p^\mathrm{rad}(t,\cdot)$, $p^\mathrm{tan}(t,\cdot)$,  have compact support. Given any smooth function $\chi(t,r)$ in $W^{1,\infty}_{\mathrm{loc}}$ and any solution of~\eqref{EVsystem} define
\[
\mathcal{I}(t)=\int_{\R^3}\chi\,q(t,r)\,dx\:.
\] 
Then the following integral identity is verified:
\begin{equation}\label{identidad_genericaII}
 \frac{d\mathcal{I}}{dt}= \int_{\R^3} \left[ e^{\mu- \lambda}p^\mathrm{rad} \frac{\partial \chi}{\partial r} - e^{\mu- \lambda}\chi \left(h\mu_r + p^\mathrm{rad} \lambda_r - \frac{p^\mathrm{tan}}{r}\right) + q \left(\frac{\partial \chi}{\partial t} - 2 \chi \lambda_t\right) \right]\, dx\:.
\end{equation}
\end{Lemma}
\begin{proof}
In~\eqref{divergence_theorem} we use
\begin{align*}
\xi_0 &=0\:, \\
\xi_i & = \chi(t,r) \frac{x_i}{r}\:.
\end{align*}
After a long but straightforward computation we obtain
\begin{equation*}
T^{\mu \nu}\nabla_\mu \xi_\nu=e^{-2 \lambda} p^\mathrm{rad} \frac{\partial \chi}{\partial r} + e^{-\lambda - \mu}q \frac{\partial \chi}{\partial t} - \chi \left[e^{-2 \lambda}h \mu_r + 2 q \lambda_t e^{-\lambda - \mu} + e^{-2 \lambda}p^\mathrm{rad} \lambda_r - p^\mathrm{tan}\frac{ e^{-2 \lambda}}{r}\right]\:.
\end{equation*}
We will choose $\Omega$ to be the coordinate image of a cylinder $[0,T]\times B(R)$. In this fashion, we have that
\begin{align}
\label{identidad_genericaI}
 \int_\Omega T^{\mu \nu}\nabla_\mu \xi_\nu \, dg = \int_{0}^{T}\int_{|x|\leq R}&\left[ e^{\mu- \lambda}p^\mathrm{rad} \frac{\partial \chi}{\partial r} - e^{\mu- \lambda}\chi \left(h\mu_r + p^\mathrm{rad} \lambda_r - \frac{p^\mathrm{tan}}{r}\right) \right.\nonumber \\ 
 &+  \left. q \left(\frac{\partial \chi}{\partial t} - 2 \chi \lambda_t\right)\right] \, dx \, dt.
\end{align}
Now we compute the corresponding boundary integral in (\ref{divergence_theorem}). First, the current reads
\begin{align*}
J^0&=q\chi e^{-\lambda - \mu}\:, \\
J^a&=e^{-2\lambda} p^\mathrm{rad} \chi(r) \frac{x^a}{r}\:.
\end{align*}
Next we write $\partial\Omega=A_1\cup A_2\cup A_3$, where
\begin{itemize}

\item $A_1=\{t=T,|x|\le R\}$. The outer unit normal is $e^{\mu(r,T)} dt$; the induced metric is $e^{2\lambda(r,T)}dr^2+r^2d\omega^2$, the volume element $e^{2 \lambda(r,T)}\,dx$.

\item $A_2=\{t=0,|x|\le R\}$. The outer unit normal is $-e^{\mu(r,0)} dt$; the induced metric is $e^{2\lambda(r,0)}dr^2+r^2d\omega^2$, the volume element $e^{2 \lambda(r,0)}\,dx$.

\item $A_3=\{0 < t < T, \, |x|=R \}$. The outer unit normal  has the form $- e^{\lambda(t,R)} \frac{x^i}{R} dx_i$. The metric is $ds^2=-e^{2\mu(t,R)}dt^2 +R^2d\omega^2$, the volume element $e^{2 \mu(t,R)}dS_Rdt$, where $dS_R$ is the surface element on the sphere of radius $R$. 

\end{itemize}

Summing up we get  
\begin{align}
 \int_{\partial \Omega} J^\mu \eta_\mu \, d\sigma_g =& \int_{|x|\leq R} q(T,r) \chi(T,r) \,dx
  - \int_{|x|\leq R} q(0,r) \chi(0,r) \,dx \nonumber \\
  &- \int_{0}^{T}\int_{|x|=R} p^\mathrm{tan}(t,R)\chi(t,R) e^{\mu(t,R)-\lambda(t,R)}\, dS_R\,dt\:.
\end{align}
Having assumed that the matter quantities are compactly supported in the variable $r$, the boundary integral vanishes in the limit $R\to\infty$, whereas the other integrals remain bounded\footnote{Of course the compact support condition can be replaced by a suitable decay assumption.}. Thus in the limit we obtain 
\[
\left[\int_{\R^3}q\chi\, dx\right]_0^T
 =\int_0^T \int_{\R^3} \left[e^{\mu- \lambda}p^\mathrm{rad} \frac{\partial \chi}{\partial r} - e^{\mu- \lambda}\chi \left(h\mu_r + p^\mathrm{rad} \lambda_r - \frac{p^\mathrm{tan}}{r}\right) + q \left(\frac{\partial \chi}{\partial t} - 2 \chi \lambda_t\right)\right] dx\,dt\:,
\]
which is the integral version of~\eqref{identidad_genericaII}.
\end{proof}

We shall now derive two particular cases of the identity~\eqref{identidad_genericaII}. First 
let us choose
\[
  \chi= e^{2 \lambda} F(r)\:,
\]
for a smooth function $F$. We have $\partial_t \chi = 2 \lambda_t e^{2 \lambda} F(r)$
and then $\partial_t \chi - 2 \chi \lambda_t=0$.
In this way equation (\ref{identidad_genericaII}) implies 
\begin{equation}
\label{identidad_genericaIII}
\left[\int_{\R^3}qe^{2 \lambda}F\, dx\right]_0^T
 =\int_0^T \int_{\R^3} e^{\mu+ \lambda} \left[p^\mathrm{rad} F'  + F\left(\lambda_r p^\mathrm{rad} - h \mu_r + \frac{p^\mathrm{tan}}{r}\right)\right] dx\,dt\:.
\end{equation}
Note now that, using (\ref{auxiliary}), (\ref{einstein2}) and (\ref{quasilocalmass}),
\begin{align*}
    \lambda_r p^\mathrm{rad} - h \mu_r & = p^\mathrm{rad}(\lambda_r + \mu_r) - \mu_r (p^\mathrm{rad} + h) = (\lambda_r + \mu_r) \left(p^\mathrm{rad} - \frac{\mu_r e^{-2\lambda}}{4\pi r} \right) \\
%
 & =  -\frac{m}{4 \pi r^3} (\lambda_r + \mu_r).
\end{align*}
Then (\ref{identidad_genericaIII}) becomes
\begin{equation}
\label{identidad_genericaV}
\left[\int_{\R^3}qe^{2 \lambda}F\, dx\right]_0^T= \int_0^T\int_{\R^3} e^{\mu +  \lambda} \left[p^\mathrm{rad} F' + p^\mathrm{tan} \frac{F}{r} - \frac{F}{r}\frac{m (\lambda_r + \mu_r)}{4 \pi r^2} \right] dx\,dt\:.
\end{equation}
In the integral in the right hand side we use that
$$
  -\int_0^T\int_{\R^3} e^{\mu +  \lambda} \frac{F}{r}\frac{ (\lambda_r + \mu_r) m}{4 \pi r^2} \, dxdt = -\int_0^T \int_0^{\infty} \frac{d e^{\lambda + \mu}}{dr}\frac{F}{r}m \, dr\,dt
$$
$$
   = \int_0^T \int_0^{\infty} \frac{d}{dr}\left( \frac{Fm}{r}\right) e^{\lambda + \mu} \,drdt - HT \left(\lim_{r \to\infty} \frac{F(r)}{r}\right).$$
This leads to 
\begin{align}
\label{identidad_genericaIV}
\left[\int_{\R^3}qe^{2 \lambda}F\, dx\right]_0^T& = - HT \left(\lim_{r \to\infty} \frac{F(r)}{r}\right)\nonumber\\
& + \int_0^T \int_{\R^3} e^{\lambda + \mu} \left[p^\mathrm{rad} F'+ p^\mathrm{tan} \frac{F}{r} + h \frac{F}{r} + \frac{m}{4 \pi r^2} \frac{d}{dr}\left(\frac{F}{r} \right) \right] \, dx\,dt\:.
\end{align}
Finally for $F(r) = r$ we obtain
\begin{equation}\label{mainformula}
\left[\int_{\R^3}qe^{2 \lambda}r\, dx\right]_0^T = - HT + \int_0^T \int_{\R^3} e^{\lambda + \mu} \left(p^\mathrm{rad}+p^\mathrm{tan}+h \right) \, dx\,dt\:.
\end{equation}
%

\subsection{Virial inequalities for steady states}
The existence of steady states solutions to the  Einstein-Vlasov system is well understood, we refer to \cite{AnRe2007,FHU} and the references therein. The identity~\eqref{mainformula} restricted to steady states imply
\begin{equation}\label{mainformulastatic}
H=\int_{\R^3}e^{\lambda+\mu}(p^\mathrm{tan}+p^\mathrm{rad}+h)\,dx\:.
\end{equation}
\begin{Remark}\textnormal{
The fundamental identity~\eqref{mainformulastatic} can be proved directly using the Einstein equations for static spherically symmetric spacetimes, see~\cite{Hakan}. Our derivation has two advantages. Firstly, we obtained~\eqref{mainformulastatic} as a special case of a more general identity which holds for time dependent solutions, see Lemma~\ref{integralidentityEV}. Secondly, the technique of the vector fields multipliers, which we used to derive~\eqref{mainformulastatic}, can also be used on spacetimes which are not spherically symmetric and therefore our argument could be useful to prove generalizations of~\eqref{mainformulastatic} for solutions with less symmetry.}
\end{Remark}
This identity leads naturally to a bound on the {\it central redshift} 
\[
Z_c=e^{-\mu(0)}-1\in [0,+\infty)
\]
in terms of the mass-energy of the static solution. We consider only static solutions of the spherically symmetric Einstein-Vlasov system.

\begin{Proposition}
\label{result}
Let $f$ be a static solution of the spherically symmetric Einstein-Vlasov system with compact support. 
Then the following inequality holds true
\begin{equation}
\label{boundredshift}
e^{\mu(0)}\leq \left\{ \begin{array}{ll} \frac{H}{M} & \mbox{ if } H \leq M \\  & \\ \frac{H}{2H -M} & \mbox{ if }H\geq M \end{array} \right. 
\quad\text{i.e }\quad
 Z_c\geq \left| \frac{M}{H}-1 \right|.
\end{equation}
\end{Proposition}
\begin{proof}
Since $\mu$ is increasing, $\mu(r)\geq\mu(0)$ and so 
\[
\int_{\R^3}e^{\lambda+\mu}(p^\mathrm{rad}+p^\mathrm{tan}+h)\geq e^{\mu(0)}\int_{\R^3}e^\lambda h\geq Me^{\mu(0)}\:.
\]
Using this in~\eqref{mainformulastatic} gives 
\begin{equation}\label{firstest}
e^{\mu(0)}\leq\frac{H}{M}\:.
\end{equation}
Moreover 
\begin{align*}
p^\mathrm{rad}+p^\mathrm{tan} + h&=h+\int_{\R^3}
\left(\frac{x\cdot
v}{r}\right)^2f\frac{dv}{\sqrt{1+|v|^2}}+\int_{\R^3}\left|\frac{x\wedge
v}{r}\right|^2f\frac{dv}{\sqrt{1+|v|^2}}\\
&= 2h+\int_{\R^3}\left(\frac{|v|^2}{\sqrt{1+|v|^2}}-\sqrt{1+|v|^2}\right)f\,dv=2h-\int_{\R^3} f\,\frac{dv}{\sqrt{1+|v|^2}}\\
\end{align*}
Thus, since $\lambda+\mu\geqslant\mu(0)$ and $e^\mu\leq 1\leq e^\lambda$,
\[
\int_{\R^3} e^{\lambda+\mu}(p^\mathrm{rad}+p^\mathrm{tan}+h)\,dx\geq e^{\mu(0)}(2H-M)
\]
and so by~\eqref{mainformulastatic},
\begin{equation}\label{secondest}
e^{\mu(0)}\leq \frac{H}{2H -M}\:, \quad \mbox {when } \quad  \frac{H}{M} > \frac12\: .
\end{equation}
The result follows from~\eqref{firstest} and~\eqref{secondest} and taking into account that  
$$
\frac{H}{M} \leq \frac{H}{2H -M}
$$ 
is satisfied in the case $\frac12 < \frac{H}{M} \leq 1$.
\end{proof}


\subsection{Shells and Jeans' type steady states}

Let $R$ be the radius support of the steady state. Using that $\mu$ is negative and increasing and that the steady state matches the Schwarzschild solution at $r=R$ we obtain the bound
\begin{equation}\label{easybound}
e^{\mu(0)}\leq e^{\mu(R)}=\sqrt{1-\frac{2H}{R}}\:.
\end{equation}
The inequality~\eqref{easybound} can be combined with~\eqref{boundredshift} to obtain an upper bound on $e^{\mu(0)}$
in terms of $R$, $H$ and $M$.

\medskip

Now, we consider briefly an important class of steady states, namely the Jeans type steady states, see \cite{W2}. For these steady states the distribution function $f$ has the form
\begin{equation}\label{jeans}
f(x,v)=\psi(E,F)\:,\text{ where }E=e^{\mu}\sqrt{1+|v|^2}\:,\ F=|x\wedge v|^2\:.
\end{equation}
Since the particles energy $E$ and the angular momentum $F$ are conserved quantities, the particle density~\eqref{jeans} is automatically a solution of the Vlasov equation. The existence of Jeans type steady states is then obtained by replacing the ansatz $f=\psi(E,F)$ into the (time independent) Einstein equations and proving existence of global solutions for the resulting system of ODEs.  We refer to~\cite{Rein2000} where this procedure is carried out for a large class of profiles $\psi$; moreover the Jeans type steady states constructed in~\cite{Rein2000} all have compact support and satisfy that
\begin{equation}\label{e0}
\exists\, E_0\in (0,1) \text{ such that }\psi=0\:, \text{ for }E\geq E_0\:. 
\end{equation}
Thus $E_0$ is the maximum particle energy in the ensemble. The property~\eqref{e0} is necessary in order that the distribution function~\eqref{jeans} be asymptotically flat and with finite energy. For Jeans type steady states one obtains a new estimate on $e^{\mu(0)}$ in a straightforward way:
\[
H=\int_{\R^3}\int_{\R^3}\sqrt{1+|v|^2}f\,dv\,dx=\int_{\R^3}\int_{\R^3}e^{-\lambda-\mu}e^\mu\sqrt{1+|v|^2}e^\lambda f\,dv\,dx\leq \frac{E_0}{e^{\mu(0)}}M\:,
\]  
whence
\begin{equation}\label{boundjeans}
e^{\mu(0)}\leq E_0\frac{M}{H}\:.
\end{equation}
{ In fact, for Jeans' type solutions we have \cite{Rein2000}
$$
   E_0 = \sqrt{1-\frac{2H}{ R}}
$$
and thus, combining \eqref{boundjeans} with \eqref{easybound} we conclude that for Jeans' type steady states  the inequality
$$
  e^{\mu(0)}\leq e^{\mu(R)}= \min \left\{1,\frac{M}{H}Ê\right\} \sqrt{1-\frac{2H}{ R}} 
$$
hods.}

Consider now the case of a static shell. Let $f$ be a static shell solution of the spherically symmetric Einstein-Vlasov system with inner radius $R_1$ and outer radius $R_2$.  Using (\ref{einstein2}) we can write $\mu(0)$ as follows
\begin{align*}
   \mu(0) 
& = - \int_0^\infty e^{2 \lambda}\left(\frac{m}{r^2} +  4\pi r p^\mathrm{rad} \right) dr \\
      & = - \int_0^\infty \frac{1}{1- 2m /r} \left(\frac{m}{r^2} +  4\pi r p^\mathrm{rad} \right) dr\: .
\end{align*}
By Buchdahl's inequality (\ref{BOU}), the identity (\ref{quasilocalmass}) and the bound $p^\mathrm{rad}\leq h$, we obtain
 \begin{align*}
   \mu(0) 
     & \ge -9 \int_{R_1}^\infty \left(\frac{m}{r^2} +  4\pi r p^\mathrm{rad} \right)  dr \\
      & \ge - 9 \int_{R_1}^\infty  \left(\frac{H}{r^2} +  4\pi r h\right) dr \\
      & \ge -\frac{9H}{R_1}- \frac{9}{R_1} \int_{R_1}^\infty 4\pi r^2 h\, dr \\
      & = -\frac{18H}{ R_1}.
\end{align*}
Now, we use the upper estimates on $\mu(0)$ of the  Proposition \ref{result} to find
$$
\mu(0) = \ln \left(\frac{1}{Z_c +1}\right) \leq \ln \left(\frac{1}{\left| \frac{M}{H}-1 \right|  +1}\right).
$$
Combining both estimates we obtain$$
R_1\leq \frac{18H}{\ln \left({\left| \frac{M}{H}-1 \right|  +1}\right)}\:,
$$
i.e., the inner radius of a static shell with given ADM energy and rest mass cannot be arbitrarily large.




\end{document}